\documentclass{amsart}

\usepackage{amsthm,amsfonts,amsmath,amssymb,mathrsfs,verbatim,cite}
\usepackage{pstricks,graphicx}

\newcommand{\D}{\textup{D}}
\newcommand{\abs}[1]{\lvert#1\rvert}
\newcommand{\Complex}{\mathbb C}
\newcommand{\Z}{\mathbb Z}
\newcommand{\la}{\lambda}
\newcommand{\qbin}[2]{\genfrac{[}{]}{0pt}{}{#1}{#2}}
\newcommand{\Mod}[1]{\!\!\!\pod{#1}}
\allowdisplaybreaks

\begin{document}

\title[The Flohr--Grabov--Koehn conjectures]
{Proof of the Flohr--Grabow--Koehn conjectures
for characters of logarithmic conformal field theory}

\author{S. Ole Warnaar}\thanks{Work supported by the Australian Research Council}
\address{Department of Mathematics and Statistics,
The University of Melbourne, VIC 3010, Australia}
\email{warnaar@ms.unimelb.edu.au}



\begin{abstract}
In a recent paper Flohr, Grabow and Koehn conjectured that the characters 
of the logarithmic conformal field theory $c_{k,1}$, of central charge
$c=1-6(k-1)^2/k$, admit fermionic representations labelled by the 
Lie algebra $\D_k$. In this note we provide a simple
analytic proof of this conjecture.
\end{abstract}

\maketitle

\section{Introduction}
In the past two decades an almost complete understanding of the
analytic and combinatorial structure of fermionic character
representations for the minimal unitary models $M(p,p')$
in conformal field theory has been obtained
\cite{Berkovich94,BM96,BMS98,DKKMM93,FQ95,FQ96,FLW97,FLPW00,Melzer94,W96a,W96b,W97,Welsh05}.
In a recent paper Flohr, Grabow and Koehn (FGK) \cite{FGK06}
took the first tentative steps towards
extending these results to the realm of logarithmic conformal 
field theory (LCFT). FGK focused on one of the simplest examples of a LCFT; 
the $\mathcal{W}(2,2k-1,2k-1,2k-1)$ series of triplet algebras
of central charge
\[
c=1-\frac{6(k-1)^2}{k},
\]
denoted as the $c_{k,1}$ models for short.
Surprisingly, FGK conjectured that the
characters of the $c_{k,1}$ models may be described in terms
of the Lie algebra $\D_k$.
For example, if $B$ denotes the inverse Cartan matrix of $\D_k$, 
then, conjecturally,
\begin{equation}\label{FGK}
\chi_{\la}(\tau)=q^{\varphi_{\la}}
\sum_{\substack{n_1,\dots,n_k=0 \\ n_{k-1}\equiv n_k\Mod{2}}}^{\infty}
\frac{q^{\sum_{i,j=1}^k B_{ij} n_i n_j 
+\frac{\la}{2}(n_{k-1}-n_k)}}{(q;q)_{n_1}\cdots(q;q)_{n_k}}.
\end{equation}
Here $\chi_{\la}$ for $\la\in\{0,1,\dots,k\}$ is a 
$c_{k,1}$ character, $\varphi_{\la}=\la^2/(4k)-1/24$,
$q=\exp(2\pi \textup{i}\tau)$,
$(q;q)_n=(1-q)(1-q^2)\cdots(1-q^n)$, and $a\equiv b\pod{c}$ is shorthand for
$a\equiv b\pmod{c}$.

In support of \eqref{FGK} FGK provided a proof 
for the degenerate case $k=2$ (in which case $B$ is half the $2\times 2$ 
identity matrix), and showed that in the $q\to 1^{-}$ limit \eqref{FGK}
gives rise to the well-known $\D_k$ dilogarithm identity
\[
2L\Bigl(\frac{1}{k}\Bigr)+\sum_{j=2}^{k-1} L\Bigl(\frac{1}{j^2}\Bigr)
=\frac{\pi^2}{6},
\]
where $L(x)$ is the Rogers dilogarithm \cite{Lewin58}.
(The $\D_k$ nature of the above identity lies with the fact that 
$x=(1/4,1/9,\dots,1/(k-1)^2,1/k,1/k)$ solves the simultaneous
equations $x_i=\prod_{j=1}^k (1-x_j)^{2B_{ij}}$ for $1\leq i\leq k$.)

\medskip

In this paper we provide an analytic
proof of \eqref{FGK} and its allied $c_{k,1}$
character identities. Key is the observation that the
matrix $B$ contains the submatrix
\[T=(\min(i,j))_{1\leq i,j\leq k-2}\]
which itself admits character identities similar to
\eqref{FGK}. Indeed it is a classical result --- first discovered
by Andrews \cite{Andrews74} in the context of partition theory ---
that
\begin{equation}\label{Andrews}
\chi^{\textup{Vir}}_{\la}(\tau)=
q^{\phi_{\la}}
\sum_{n_1,\dots,n_{k-2}=0}^{\infty}
\frac{q^{\sum_{i,j=1}^{k-2} T_{ij} n_i n_j+\sum_{i=\la}^{k-2} (i-\la+1)n_i}}
{(q;q)_{n_1}\cdots(q;q)_{n_{k-2}}},
\end{equation}
where $\chi^{\textup{Vir}}_{\la}$ for $\la\in\{1,\dots,k-1\}$
are the characters of the Virasoro
minimal models $M(2,2k-1)$ and 
\[
\phi_{\la}=\frac{(2\la-2k+1)^2}{8(2k-1)}-\frac{1}{24}.
\]
It is the connection between the conjectural \eqref{FGK}
and Andrews' \eqref{Andrews} that will play a crucial role in our proof.

\section{$c_{k,1}$ character formulas}

We will not formally define the characters of the $c_{k,1}$ models
but simply state their bosonic representation as obtained in 
\cite{Flohr96}.

Throughout we let $\tau\in\Complex$, $\text{Im}(\tau)>0$ and $q\in\Complex$
be related by $q=\exp(2\pi \textup{i}\tau)$, so that $\abs{q}<1$.
This will make all $q$-series considered in this paper absolutely convergent
so that we need not concern ourselves with order of summation in multiple
series. We also use the standard $q$-notations
\[
(a;q)_n=\prod_{i=0}^{n-1} (1-aq^i) \qquad\text{and}
\qquad (a;q)_{\infty}=\prod_{i=0}^{\infty}(1-aq^i),
\]
and Dedekind's eta-function
\[
\eta(\tau)=q^{1/24}(q;q)_{\infty}.
\]
Finally we need the theta and affine theta functions
\begin{align*}
\Theta_{\la,k}(\tau)&=\sum_{n\in\Z+\frac{\la}{2k}}
q^{kn^2}, \\
(\partial\Theta)_{\la,k}(\tau)&=\sum_{n\in\Z+\frac{\la}{2k}}
2kn\, q^{k n^2}.
\end{align*}

With the above notation the following bosonic character formulas
corresponding to the $c_{k,1}$ LCFT hold \cite{Flohr96}:
\begin{align*}
\chi_{\la}(\tau)&=\frac{\Theta_{\lambda,k}(\tau)}{\eta(\tau)},\\[2mm]
\chi_{\la}^{+}(\tau)&=\frac{(k-\la)\,\Theta_{\la,k}(\tau)+
(\partial\Theta)_{\la,k}(\tau)}{k\,\eta(\tau)}=
\frac{1}{\eta(\tau)}
\sum_{n\in\Z}(2n+1) q^{k(n+\frac{\la}{2k})^2},\\[2mm]
\chi_{\la}^{-}(\tau)&=\frac{\la\,\Theta_{\la,k}(\tau)-
(\partial\Theta)_{\la,k}(\tau)}{k\,\eta(\tau)}=
\frac{1}{\eta(\tau)}
\sum_{n\in\Z}2n\, q^{k(n-\frac{\la}{2k})^2},
\end{align*}
with $\la\in\{0,1,\dots,k\}$ in $\chi_{\la}$ and
$\la\in\{1,\dots,k-1\}$ in $\chi_{\la}^{\pm}$.

To state the fermionic expressions of FGK we let
$B$ be the inverse Cartan matrix of the 
Lie algebra $\D_k$ with labelling of the vertices of the Dynkin
diagram given by

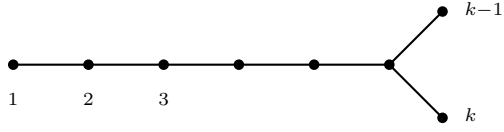
\mbox{}\bigskip\bigskip

\begin{center}
\psset{unit=1cm}
\begin{pspicture}(6,0)(0,0)
\psline(0,0)(5,0)
\psline(5,0)(5.7071,0.7071)
\psline(5,0)(5.7071,-0.7071)
\pscircle*(0,0){0.07}
\pscircle*(1,0){0.07}
\pscircle*(2,0){0.07}
\pscircle*(3,0){0.07}
\pscircle*(4,0){0.07}
\pscircle*(5,0){0.07}
\pscircle*(5.7071,0.7071){0.07}
\pscircle*(5.7071,-0.7071){0.07}
\uput[-90](0,-0.2){$\scriptstyle 1$}
\uput[-90](1,-0.2){$\scriptstyle 2$}
\uput[-90](2,-0.2){$\scriptstyle 3$}
\put(6,0.66){$\scriptstyle k-1$}
\put(6,-0.75){$\scriptstyle k$}
\end{pspicture}
\end{center}

\bigskip\bigskip\bigskip

\noindent
Hence 
\begin{subequations}\label{B}
\begin{gather}
B_{k,k-1}=B_{k-1,k}=\frac{k-2}{4},\qquad 
B_{k-1,k-1}=B_{k,k}=\frac{k}{4}, \\
B_{i,k-1}=B_{i,k}=B_{k-1,i}=B_{k,i}=\frac{i}{2} \qquad 1\leq i\leq k-2, \\[2mm]
B_{ij}=\min(i,j)\qquad  1\leq i,j,\leq k-2.
\end{gather}
\end{subequations}
Then the conjectures of FGK correspond to
\begin{align*}
\chi_{\la}(\tau)&=q^{\varphi_{\la}}
\sum_{\substack{n_1,\dots,n_k=0 \\ n_{k-1}\equiv n_k\Mod{2}}}^{\infty}
\frac{q^{\sum_{i,j=1}^k B_{ij} n_i n_j 
+\frac{\la}{2}(n_{k-1}-n_k)}}{(q;q)_{n_1}\cdots(q;q)_{n_k}}, \\
\chi_{\la}^{+}(\tau)&=q^{\varphi_{\la}}
\sum_{\substack{n_1,\dots,n_k=0 \\ n_{k-1}\equiv n_k\Mod{2}}}^{\infty}
\frac{q^{\sum_{i,j=1}^k B_{ij} n_i n_j 
+\sum_{i=k-\la}^{k-2} (i-k+\la+1)n_i
+\frac{\la}{2}(n_{k-1}+n_k)}}{(q;q)_{n_1}\cdots(q;q)_{n_k}} \\
\intertext{and}
\chi_{k-\la}^{-}(\tau)&=q^{\varphi_{\la}} 
\sum_{\substack{n_1,\dots,n_k=0 \\ n_{k-1}\not\equiv n_k\Mod{2}}}^{\infty}
\frac{q^{\sum_{i,j=1}^k B_{ij} n_i n_j 
+\sum_{i=k-\la}^{k-2} (i-k+\la+1)n_i
+\frac{\la}{2}(n_{k-1}+n_k)}}{(q;q)_{n_1}\cdots(q;q)_{n_k}},
\end{align*}
where
\[
\varphi_{\la}=\frac{\la^2}{4k}-\frac{1}{24}.
\]
Note that the last two expressions have identical summand
and differ only in the restriction on the parity of
$n_{k-1}+n_k$.

In addition to the above three conjectures we will also prove that
\[
\chi_{k-\la}(\tau)=q^{\varphi_{\la}}
\sum_{\substack{n_1,\dots,n_k=0 \\ n_{k-1}\not\equiv n_k\Mod{2}}}^{\infty}
\frac{q^{\sum_{i,j=1}^k B_{ij} n_i n_j 
+\frac{\la}{2}(n_{k-1}-n_k)}}{(q;q)_{n_1}\cdots(q;q)_{n_k}}
\]
so that we have two fermionic representations for every character $\chi_{\la}$.

Equating each of the fermionic forms with the corresponding bosonic form
we obtain the following two $q$-series identities:
\begin{subequations}
\begin{multline}\label{BF1}
\sum_{\substack{n_1,\dots,n_k=0 \\ n_{k-1}+n_k\equiv \sigma\Mod{2}}}^{\infty}
\frac{q^{\sum_{i,j=1}^k B_{ij} n_i n_j 
+\frac{\la}{2}(n_{k-1}-n_k+\sigma)
-\frac{1}{4}\sigma k}}
{(q;q)_{n_1}\cdots(q;q)_{n_k}} \\
=\frac{1}{(q;q)_{\infty}} \sum_{n=-\infty}^{\infty} q^{kn^2+(\la-\sigma k)n}
\end{multline}
for $\la\in\{0,\dots,k\}$ and
\begin{multline}\label{BF2}
\sum_{\substack{n_1,\dots,n_k=0 \\ n_{k-1}+n_k\equiv \sigma\Mod{2}}}^{\infty}
\frac{q^{\sum_{i,j=1}^k B_{ij} n_i n_j 
+\sum_{i=k-\la}^{k-2} (i-k+\la+1)n_i
+\frac{\la}{2}(n_{k-1}+n_k+\sigma)-\frac{1}{4}\sigma k}}
{(q;q)_{n_1}\cdots(q;q)_{n_k}} \\
=\frac{1}{(q;q)_{\infty}}\sum_{n=-\infty}^{\infty}
(2n-\sigma+1)q^{kn^2+(\la-\sigma k) n}
\end{multline}
for $\la\in\{1,\dots,k-1\}$. In both formulas $\sigma$ is either zero or one.
\end{subequations}

We remark that by the Jacobi triple product identity \cite{Andrews76}
the right-hand side of \eqref{BF1} may also be written in product form
as
\[
\frac{(-q^{k+\la-\sigma k},-q^{k-\la+\sigma k},q^{2k};q^{2k})_{\infty}}
{(q;q)_{\infty}},
\]
where 
\[
(a,q/a,q;q)_{\infty}=\prod_{i=1}^{\infty}(1-a q^{i-1})(1-q^i/a)(1-q^i).
\]

\section{Proof of \eqref{BF1} and \eqref{BF2}}\label{SecProof}

As a first step we rewrite \eqref{BF1} and \eqref{BF2} by 
replacing the summation variables $n_{k-1}$ and $n_k$ by $n$ and $m$ 
respectively. Also eliminating explicit reference to the inverse
Cartan matrix $B$ using \eqref{B}, we get
\begin{subequations}
\begin{multline}\label{id1}
\sum_{\substack{n,m=0 \\ n+m\equiv\sigma\Mod{2}}}^{\infty}
\frac{q^{\frac{k}{4}(n^2+m^2-\sigma)+\frac{k-2}{2} nm
+\frac{\la}{2}(n-m+\sigma)}} {(q;q)_n(q;q)_m} \\
\times
\sum_{n_1,\dots,n_{k-2}=0}^{\infty}
\frac{q^{N_1^2+\cdots+N_{k-2}^2+(n+m)(N_1+\cdots+N_{k-2})}}
{(q;q)_{n_1}\cdots (q;q)_{n_{k-2}}} \\
=\frac{1}{(q;q)_{\infty}}\sum_{n=-\infty}^{\infty}
q^{kn^2+(\la-\sigma k)n} \qquad\qquad\qquad
\end{multline}
and
\begin{multline}\label{id2}
\sum_{\substack{n,m=0 \\ n+m\equiv\sigma\Mod{2}}}^{\infty}
\frac{q^{\frac{k}{4}(n^2+m^2-\sigma)+\frac{k-2}{2}nm
+\frac{\la}{2}(n+m+\sigma)}}{(q;q)_n(q;q)_m} \\
\times \sum_{n_1,\dots,n_{k-2}=0}^{\infty}
\frac{q^{N_1^2+\cdots+N_{k-2}^2+N_{k-\la}+\cdots+N_{k-2}
+(n+m)(N_1+\cdots+N_{k-2})}}{(q;q)_{n_1}\cdots (q;q)_{n_{k-2}}} \\
=\frac{1}{(q;q)_{\infty}}\sum_{n=-\infty}^{\infty}
(2n-\sigma+1)q^{kn^2+(\la-\sigma k)n}, \qquad
\end{multline}
\end{subequations}
where $N_i=n_i+n_{i+1}+\cdots+n_{k-2}$.
We note that the quadratic form involving the $N_i$ 
may alternatively be expressed in terms of the 
submatrix $T$ of $B$ given by $T_{ij}=\min(i,j)$ for $1\leq i,j\leq k-2$.
Specifically,
\[
N_1^2+\cdots+N_{k-2}^2
=\sum_{i,j=1}^{k-2} T_{ij} n_i n_j.
\]

\begin{proof}[Proof of \eqref{id1}]
To prove \eqref{id1} we denote its left-hand side by $L_{\la,k,\sigma}$.
Shifting the summation index $n\to 2n-m-\sigma$ and replacing $k\to k+1$
we obtain
\begin{multline}\label{L}
L_{\la,k+1,\sigma}=\sum_{n=0}^{\infty}\sum_{m=0}^{2n-\sigma}
\frac{q^{kn^2+(n-m)(n-m+\la-\sigma)-\sigma k n}}
{(q;q)_{2n-m-\sigma}(q;q)_m} \\
\times
\sum_{n_1,\dots,n_{k-1}}^{\infty}
\frac{q^{N_1^2+\cdots+N_{k-1}^2+(2n-\sigma)(N_1+\cdots+N_{k-1})}}
{(q;q)_{n_1}\cdots (q;q)_{n_{k-1}}},
\end{multline}
where now 
\[
N_i=n_i+\cdots+n_{k-1}
\]
and $\la\in\{0,\dots,k+1\}$.

In order the evaluate the multisum on the second line we consider
the more general expression
\begin{equation}\label{An}
Q_{k,i}(x)=
\sum_{n_1,\dots,n_{k-1}=0}^{\infty} 
\frac{x^{N_1+\cdots+N_{k-1}}q^{N_1^2+\cdots+N_{k-1}^2+N_i+\cdots+N_{k-1}}}
{(q;q)_{n_1}\cdots (q;q)_{n_{k-1}}}
\end{equation}
for $i\in\{1,\dots,k\}$.
This multisum was first introduced by Andrews
\cite[Equation (2.5)]{Andrews74} in his proof of the 
analytic form of Gordon's partition identities.
(In the notation of Andrews' book \textit{Partition Theory} we have
$Q_{k,i}(x)=J_{k,i}(0;x;q)$, see  \cite[Equation (7.3.8)]{Andrews76}.)

In \cite[Equation (2.1)]{Andrews74} 
(see also \cite[Equations (7.2.1) \& (7.2.2)]{Andrews76})
we find the following single-sum form for $Q_{k,i}$:
\[
Q_{k,i}(x)=\frac{1}{(xq;q)_{\infty}}
\sum_{j=0}^{\infty} (-1)^j x^{kj} q^{\binom{j}{2}+
kj^2+(k-i+1)j}(1-x^i q^{i(2j+1)})\;\frac{(xq;q)_j}{(q;q)_j}.
\]
This in fact shows that $Q_{k,i}$ coincides with functions
studied earlier by Rogers \cite{Rogers17} and Selberg 
\cite{Selberg36}. 
From the above we infer that
\begin{multline}\label{Kkla}
Q_{k,k-\la+1}(q^{2n-\sigma})=\frac{1}{(q;q)_{\infty}}
\sum_{j=0}^{\infty} (-1)^j q^{\binom{j}{2}+
kj^2+(\la-\sigma k)j+2knj} \\ \times (1-q^{(k-\la+1)(2j+2n-\sigma+1)})\;
\frac{(q;q)_{j+2n-\sigma}}{(q;q)_j}.
\end{multline}

Let us now return to \eqref{L}.
By \eqref{An} the multisum on the second line of \eqref{L}
may be identified as $Q_{k,k}(q^{2n-\sigma})$ and by \eqref{Kkla} 
with $\la=1$ this may be simplified to a single sum.
Therefore
\begin{multline*}
L_{\la,k+1,\sigma}=
\frac{1}{(q;q)_{\infty}}
\sum_{j,n=0}^{\infty}\sum_{m=0}^{2n-\sigma}
(-1)^j q^{\binom{j+1}{2}+k(j+n)^2+(n-m)(n-m+\la-\sigma)
-\sigma k (j+n)} \\ 
\times (1-q^{k(2j+2n-\sigma+1)})\;
\frac{(q;q)_{j+2n-\sigma}}{(q;q)_j(q;q)_m(q;q)_{2n-m-\sigma}}.
\end{multline*}
Our next step is to shift the summation indices $n\to n-j$ and 
$m\to m-j$, resulting in
\begin{multline*}
L_{\la,k+1,\sigma}=\frac{1}{(q;q)_{\infty}}
\sum_{n=0}^{\infty}\sum_{m=0}^{2n-\sigma}
q^{kn(n-\sigma)+(n-m)(n-m+\la-\sigma)} 
(1-q^{k(2n-\sigma+1)}) \\
\times
\sum_{j=0}^{\min(m,2n-m-\sigma)}
(-1)^j q^{\binom{j+1}{2}} \;
\frac{(q;q)_{2n-j-\sigma}}{(q;q)_j(q;q)_{m-j}(q;q)_{2n-m-j-\sigma}}.
\end{multline*}
Employing standard basic hypergeometric notation \cite{GR04}
this may also we written as
\begin{multline}\label{Lphi}
L_{\la,k+1,\sigma}=\frac{1}{(q;q)_{\infty}}
\sum_{n=0}^{\infty}\sum_{m=0}^{2n-\sigma}
q^{kn(n-\sigma)+(n-m)(n-m+\la-\sigma)} (1-q^{k(2n-\sigma+1)})\\
\times 
\frac{(q;q)_{2n-\sigma}}{(q;q)_m(q;q)_{2n-m-\sigma}} \;
{_2}\phi_1\biggl[\genfrac{}{}{0pt}{}
{q^{-m},q^{-(2n-m-\sigma)}}{q^{-(2n-\sigma)}};q,q\biggr].
\end{multline}
To proceed we need the $q$-Chu--Vandermonde sum 
\cite[Equation (II.6)]{GR04}
\[
{_2}\phi_1\biggl[\genfrac{}{}{0pt}{}
{a,q^{-n}}{c};q,q\biggr]
=a^n \frac{(c/a;q)_n}{(c;q)_n}.
\]
Hence the $_2\phi_1$ series may be summed to
\begin{equation}\label{cv}
\frac{(q;q)_m(q;q)_{2n-m-\sigma}}{(q;q)_{2n-\sigma}}
\end{equation}
leading to
\[
L_{\la,k+1,\sigma}
=\frac{1}{(q;q)_{\infty}}
\sum_{n=0}^{\infty}\sum_{m=0}^{2n-\sigma}
q^{kn(n-\sigma)+(n-m)(n-m+\la-\sigma)} (1-q^{k(2n-\sigma+1)}).
\]
The remainder of the proof requires only elementary manipulations:
\begin{align*}
L_{\la,k+1,\sigma}
&=\frac{1}{(q;q)_{\infty}}
\sum_{n=0}^{\infty}\sum_{m=-n}^{n-\sigma}
q^{kn(n-\sigma)+m(m-\la+\sigma)} (1-q^{k(2n-\sigma+1)}) \\
&=\frac{1}{(q;q)_{\infty}}
\sum_{m=-\infty}^{\infty} q^{m(m-\la+\sigma)}
\sum_{n=\max(-m,m+\sigma)}^{\infty} \Bigl(q^{kn(n-\sigma)}
-q^{k(n+1)(n+1-\sigma)}\Bigr) \\
&=\frac{1}{(q;q)_{\infty}}
\sum_{m=-\infty}^{\infty} q^{(k+1)m^2-(\la-\sigma(k+1))m}.
\end{align*}
Finally replacing $k\to k-1$ and changing the summation index
$m\to -n$ completes the proof of \eqref{id1}.
\end{proof}

\begin{proof}[Proof of \eqref{id2}]
As in the proof of \eqref{id1} we denote the
left-hand side of \eqref{id2} by $L_{\la,k,\sigma}$.
Again we carry out the shift $n\to 2n-m-\sigma$ in
the summation index, and replace $k\to k+1$. Thus
\begin{multline*}
L_{\la,k+1,\sigma}=\sum_{n=0}^{\infty}\sum_{m=0}^{2n-\sigma}
\frac{q^{kn^2+(n-m)(n-m-\sigma)+(\la-\sigma k)n}}{(q;q)_{2n-m-\sigma}(q;q)_m} \\
\times\sum_{n_1,\dots,n_{k-1}=0}^{\infty} 
\frac{q^{N_1^2+\cdots+N_{k-1}^2+N_{k-\la+1}+\cdots+N_{k-1}
+(2n-\sigma)(N_1+\cdots+N_{k-1})}}
{(q;q)_{n_1}\cdots (q;q)_{n_{k-1}}}.
\end{multline*}
{}From \eqref{An} we infer that the second line is 
$J_{k,k-\la+1}(q^{2n-\sigma})$ so that we may invoke
\eqref{Kkla} to obtain
\begin{multline*}
L_{\la,k+1,\sigma}=\frac{1}{(q;q)_{\infty}}
\sum_{j,n=0}^{\infty}\sum_{m=0}^{2n-\sigma}
(-1)^j q^{\binom{j}{2}+k(j+n)^2+(\la-\sigma k)(j+n)
+(n-m)(n-m-\sigma)}\\
\times
(1-q^{(k-\la+1)(2j+2n-\sigma+1)}) \;
\frac{(q;q)_{j+2n-\sigma}}{(q;q)_j(q;q)_m(q;q)_{2n-m-\sigma}}.
\end{multline*}
Following the earlier proof we shift $n\to n-j$ and $m\to m-j$,
and use basic hypergeometric notation to find
\begin{multline*}
L_{\la,k+1,\sigma}=\frac{1}{(q;q)_{\infty}}
\sum_{n=0}^{\infty}\sum_{m=0}^{2n-\sigma}
q^{kn^2+(\la-\sigma k)n+(n-m)(n-m-\sigma)}(1-q^{(k-\la+1)(2n-\sigma+1)})  \\
\times
\frac{(q;q)_{2n-\sigma}}{(q;q)_m(q;q)_{2n-m-\sigma}}\;
{_2}\phi_1\biggl[\genfrac{}{}{0pt}{}
{q^{-m},q^{-(2n-m-\sigma)}}{q^{-(2n-\sigma)}};q,1\biggr].
\end{multline*}
This time we need the second form of the 
$q$-Chu--Vandermonde sum \cite[Equation (II.7)]{GR04}
\begin{equation}\label{qCV}
{_2}\phi_1\biggl[\genfrac{}{}{0pt}{}
{a,q^{-n}}{c};q,\frac{cq^n}{a}\biggr]
=\frac{(c/a;q)_n}{(c;q)_n}
\end{equation}
to sum the $_2\phi_1$ series to 
\[
q^{(2n-\sigma)m-m^2}\;
\frac{(q;q)_m(q;q)_{2n-m-\sigma}}{(q;q)_{2n-\sigma}}.
\]
Hence
\begin{align}\label{last}
L_{\la,k,\sigma}&=\frac{1}{(q;q)_{\infty}}
\sum_{n=0}^{\infty}\sum_{m=0}^{2n-\sigma}
q^{kn^2+(\la-\sigma k)n} 
(1-q^{(k-\la)(2n-\sigma+1)}) \\
&=\frac{1}{(q;q)_{\infty}}
\sum_{n=0}^{\infty}(2n-\sigma+1)
q^{k n^2+(\la-\sigma k)n} 
(1-q^{(k-\la)(2n+1)}) \notag \\
&=\frac{1}{(q;q)_{\infty}}
\sum_{n=-\infty}^{\infty}(2n-\sigma+1)
q^{k n^2+(\la-\sigma k)n}, \notag
\end{align}
establishing \eqref{id2}.
\end{proof}

\section{Discussion}
The $c_{k,1}$ character identities proved in this paper
admit polynomial analogues. Defining the $q$-binomial coefficient as
\begin{equation}\label{qbinomial}
\qbin{n+m}{n}=
\frac{(q;q)_{n+m}}{(q;q)_n(q;q)_m}
\end{equation}
for $n,m$ nonnegative integers, and assuming $k\geq 3$,
we for example have
\begin{equation}\label{pol}
\sum_{\substack{n_1,\dots,n_k=0 \\ n_{k-1}\equiv n_k\Mod{2}}}^{\infty}
z^{\frac{1}{2}(n_{k-1}-n_k)} q^{\sum_{i,j=1}^k B_{ij} n_i n_j}
\prod_{i=1}^k \qbin{n_i+m_i}{n_i}
=\sum_{n=-\infty}^{\infty} z^n q^{kn^2}\qbin{2L}{L-kn}.
\end{equation}
Here the $m_i$ appearing in the $q$-binomial coefficients
are fixed by
\[
m_i=\sum_{j=1}^k B_{ij}(2L\delta_{j,1}-2n_j).
\]
When $L$ tends to infinity and $z$ is specialised to $q^{\la}$
the identity \eqref{pol} simplifies to \eqref{BF1} with $\sigma=0$.
It is interesting to note that for $q=1$ it provides an identity
for the number of walks of length $2L$ on the rooted cyclic
graph $C_{2k}$ beginning and terminating at the root.
Here the parameter $z$ in the generating function serves to keep
track of the number of cycles of the rooted walks on $C_{2k}$.

The previous method of proof fails to also deal with \eqref{pol}
but, as will be shown in Appendix~\ref{appA}, \eqref{pol}
may be proved by induction on $k$.

Finally we remark that if we replace $q\to 1/q$ in \eqref{pol} and
then let $L$ tend to infinity we obtain the dual identity
\begin{align*}
\sum_{\substack{m_1,\dots,m_k=0 \\ m_1,\dots,m_{k-2}\equiv 0\Mod{2} \\
m_{k-1}\equiv m_k\Mod{2}}}^{\infty} &
\frac{z^{\frac{1}{2}(m_{k-1}-m_k)}
q^{\frac{1}{4}\sum_{i,j=1}^k C_{ij} m_i m_j}}{(q;q)_{m_1}}
\prod_{i=2}^k \qbin{n_i+m_i}{m_i} \\
&=\frac{1}{(q;q)_{\infty}}
\sum_{n=-\infty}^{\infty} z^n q^{k(k-1)n^2} \\[2mm]
&=\frac{(-zq^{k(k-1)},-q^{k(k-1)}/z,q^{2k(k-1)};q^{2k(k-1)})_{\infty}}
{(q;q)_{\infty}},
\end{align*}
where $C=B^{-1}$ is the $\D_k$ Cartan matrix and
\[
n_i=-\frac{1}{2}\sum_{j=1}^k C_{ij}m_j.
\]

\section{Postscript}
Shortly after completing this paper
B.~Feigin, E.~Feigin and Tipunin proved another family 
of character formulas for the $c_{k,1}$ models \cite{FFT07}.
Replacing
$p\to k$ and $s\to k-\la$ and $(n_{+},n_{-})\to (n,m)$
in \cite[Theorem 1.1]{FFT07} the result of Feigin \textit{et al.}
reads
\begin{multline}\label{eqFFT}
\chi_{\la}^{+}(q)+\chi_{k-\la}^{-}(q)=q^{\varphi_{\la}}
\sum_{n,m=0}^{\infty}
\frac{q^{\frac{k}{4}(n+m)^2+\frac{\la}{2}(n+m)}}
{(q;q)_n(q;q)_m} \\
\times\sum_{n_1,\dots,n_{k-1}=0}^{\infty}
\frac{q^{N_1^2+\cdots+N_{k-1}^2+N_{k-\la}+\cdots+N_{k-1}+
(n+m)(N_1+\cdots+N_{k-1})}}
{(q;q)_{n_1}\cdots (q;q)_{n_{k-1}}},
\end{multline}
where
\[
N_i=n_1+\cdots+n_{k-1}.
\]
When the sum over $n$ and $m$ is restricted to even (odd)
values of $n+m$ we obtain $\chi_{\la}^{+}(q)$ ($\chi_{k-\la}^{-}(q)$),
and in Appendix~\ref{appB} the method used in 
Section~\ref{SecProof} to prove the
FGK conjectures is employed to establish that
\begin{multline}\label{eqB}
\sum_{\substack{n,m=0 \\ n+m\equiv\sigma\Mod{2}}}^{\infty}
\frac{q^{\frac{k}{4}((n+m)^2-\sigma))
+\frac{\la}{2}(n+m+\sigma)}}{(q;q)_n(q;q)_m} \\
\times \sum_{n_1,\dots,n_{k-1}=0}^{\infty}
\frac{q^{N_1^2+\cdots+N_{k-1}^2+N_{k-\la}+\cdots+N_{k-1}
+(n+m)(N_1+\cdots+N_{k-1})}}{(q;q)_{n_1}\cdots (q;q)_{n_{k-1}}} \\
=\frac{1}{(q;q)_{\infty}}\sum_{n=-\infty}^{\infty}
(2n-\sigma+1)q^{kn^2+(\la-\sigma k)n}.
\end{multline}
This is to be compared with \eqref{id2}.
Summing the above over $\sigma\in\{0,1\}$ yields 
\eqref{eqFFT}.

\section*{Acknowledgements}
I wish to thank Omar Foda for helpful discussions
and for pointing out the conjectures of \cite{FGK06}.

\appendix

\section{}\label{appA}
In the appendix we prove \eqref{pol}.
To begin we replace $n_k\to 2n+n_{k-1}$ on the
left and $n\to -n$ on the right.
Then equating coefficients of $z^{-n}$ and finally replacing $k\to k+1$
yields
\begin{multline*}
\sum_{n_1,\dots,n_k=0}^{\infty}
q^{\sum_{i=1}^kN_i(N_i+2n)}
\qbin{L-(k-1)n-\sum_{i=1}^{k-1}N_i}{n_k+2n}
\qbin{L-(k-1)n-\sum_{i=1}^{k-1}N_i}{n_k} \\
\times
\prod_{i=1}^{k-1} \qbin{2L-2in+n_i-2\sum_{j=1}^iN_j}{n_i}
=\qbin{2L}{L-(k+1)n},
\end{multline*}
where
\[
N_i=n_1+\cdots+n_k.
\]
Note that we may without loss of generality assume from now on that
$n$ is a nonnegative integer. Indeed, by the shift $n_k\to n_k-2n$
we obtain the same identity but with $n$ replaced by $-n$.

Next we use the symmetry in $n$ and $m$ of the $q$-binomial coefficient
\eqref{qbinomial} to rewrite the above multisum as
\begin{multline}\label{rew}
\sum_{n_1,\dots,n_k=0}^{\infty}
q^{\sum_{i=1}^kN_i(N_i+2n)}
\qbin{L-(k-1)n-\sum_{i=1}^{k-1}N_i}{L-(k+1)n-\sum_{i=1}^k N_i}
\qbin{L-(k-1)n-\sum_{i=1}^{k-1}N_i}{n_k} \\
\times
\prod_{i=1}^{k-1} \qbin{2L-2in+n_i-2\sum_{j=1}^iN_j}{n_i}
=\qbin{2L}{L-(k+1)n}.
\end{multline}
At first sight this may not appear at all significant, but a close inspection
reveals that we may now replace \eqref{qbinomial} by
\begin{equation}\label{qbinomial2}
\qbin{n+m}{n}=\begin{cases} \displaystyle
\frac{(q^{m+1};q)_n}{(q;q)_n} & \text{for $n\geq 0$} \\[2.5mm]
0 & \text{for $n<0$}.
\end{cases}
\end{equation}
The difference with the earlier definition is that the above $q$-binomial
coefficient is non-zero when $n+m<0$ and $n\geq 0$.
Clearly, if we can show that negative upper entries cannot occur in the
$q$-binomial coeffients of \eqref{rew}, then the change of definition 
is justified.
To achieve this we note that both $q$-binomial definitions imply that
the summand of \eqref{rew} vanishes unless $n_1,\dots,n_k\geq 0$
and
\[
\sum_{j=1}^k N_j\leq L-(k+1)n.
\]
But this implies that
\[
L-(k-1)n-\sum_{i=1}^{k-1}N_i\geq 2n\geq 0
\]
and
\[
2L-2in+n_i-2\sum_{j=1}^iN_j\geq 2(k-i+1)n+n_i\geq 0
\]
as required.

We now proceed by proving the identity
\begin{multline}\label{L1L2}
\sum_{n_1,\dots,n_k=0}^{\infty}
q^{\sum_{i=1}^k N_i(N_i+m)}
\qbin{L_1+m-\sum_{i=1}^{k-1}N_i}{L_1-\sum_{i=1}^kN_i}
\qbin{L_2-km-\sum_{i=1}^{k-1}N_i}{n_k} \\ \times
\prod_{i=1}^{k-1}\qbin{L_1+L_2-i m+n_i-2\sum_{j=1}^i N_j}{n_i}=
\qbin{L_1+L_2}{L_1},
\end{multline}
where $L_1,L_2,m$ are arbitrary integers.
(For $L_1<0$ both sides trivially vanish since
the sum over the $n_i$ is bounded by
$\sum_i N_i\leq L_1$.)
The identity \eqref{rew} is recovered by taking
\[
(L_1,L_2,m)\to (L-(k+1)n,L+(k+1)n,2n).
\]
Before we continue let us remark that, generally,
\eqref{L1L2} is not true if one assumes
definition \eqref{qbinomial} of the $q$-binomial coefficient.

Key to our proof of \eqref{L1L2} are 
the polynomial form of the $q$-Pfaff--Saalsch\"utz sum
\cite[Equation (3.3.11)]{Andrews76} 
\begin{equation}\label{qPS}
\sum_{n=0}^{\min(b,d)} q^{n(n+a-b)}
\qbin{a}{b-n}\qbin{c}{n}\qbin{a+c+d-n}{d-n}=
\qbin{a+d}{b}\qbin{a-b+c+d}{d}
\end{equation}
and its $d\to\infty$ limit 
(which corresponds to a polynomial analogue of the $q$-Chu--Vandermonde
sum \eqref{qCV})
\begin{equation}\label{qCVpol}
\sum_{n=0}^b q^{n(n+a-b)}
\qbin{a}{b-n}\qbin{c}{n}=\qbin{a+c}{b}.
\end{equation}
Thanks to \eqref{qbinomial2} the above two summations are
true for all integers $a,b,c,d$.

We now eliminate the variables $n_i$ by $n_i=N_i-N_{i+1}$ (with $N_{k+1}=0$)
from \eqref{L1L2} to obtain the equivalent formula
\begin{multline}\label{idN}
\sum_{N_1\geq\dots\geq N_k\geq 0}
q^{\sum_{i=1}^k N_i(N_i+m)}\qbin{L_1+m-\sum_{i=1}^{k-1} N_i}
{L_1-\sum_{i=1}^k N_i}
\qbin{L_2-km-\sum_{i=1}^{k-1} N_i}{N_k} \\ \times
\prod_{i=1}^{k-1}\qbin{L_1+L_2-i m+N_i-N_{i+1}-2\sum_{j=1}^i N_j}{N_i-N_{i+1}}=
\qbin{L_1+L_2}{L_1}.
\end{multline}
For $k=1$ this is
\[
\sum_{N_1=0}^{L_1}
q^{N_1(N_1+m)}\qbin{L_1+m}{L_1-N_1}\qbin{L_2-m}{N_1}=\qbin{L_1+L_2}{L_1},
\]
which follows \eqref{qCVpol}.
Now assume that $k\geq 2$ and write the left-hand side of \eqref{idN} as
$f_k$. Then
\begin{align*}
f_k&=\sum_{N_1\geq \dots\geq N_{k-1}\geq 0}
q^{\sum_{i=1}^{k-1}N_i(N_i+m)} 
\prod_{i=1}^{k-2}\qbin{L_1+L_2-i m+N_i-N_{i+1}-2\sum_{j=1}^i N_j}{N_i-N_{i+1}}\\
&\qquad\qquad \times 
\sum_{N_k\geq 0} q^{N_k(N_k+m)}
\qbin{L_1+m-\sum_{i=1}^{k-1} N_i}{L_1-\sum_{i=1}^k N_i} 
\qbin{L_2-km-\sum_{i=1}^{k-1} N_i}{N_k}  \\
&\qquad\qquad\qquad\qquad \times
\qbin{L_1+L_2-(k-1)m+N_{k-1}-N_k-2\sum_{j=1}^{k-1} N_j}{N_{k-1}-N_k}.
\end{align*}
The sum over $N_k$ may be performed by \eqref{qPS}, resulting in
\[
f_k=f_{k-1}.
\]
A standard induction argument completes the proof.

\section{}\label{appB}
In this appendix we prove \eqref{eqB}.
First we shift $n\to 2n-m-\sigma$ to find
\[
\text{LHS}\eqref{eqB}=
\sum_{n=0}^{\infty}\sum_{m=0}^{2n-\sigma}
\frac{q^{kn^2+(\la-\sigma k)n}}
{(q;q)_{2n-m-\sigma}(q;q)_m}\: Q_{k,k-\la}(q^{2n-\sigma}).
\]
with $Q_{k,i}(x)$ defined in \eqref{An}.
By \eqref{Kkla} with $\la\to\la+1$ this becomes
\begin{multline*}
\text{LHS}\eqref{eqB}=
\frac{1}{(q;q)_{\infty}}
\sum_{j,n=0}^{\infty}\sum_{m=0}^{2n-\sigma}
(-1)^j q^{\binom{j+1}{2}+k(j+n)^2+(\la-\sigma k)(j+n)}
\\ \times (1-q^{(k-\la)(2j+2n-\sigma+1)})\;
\frac{(q;q)_{j+2n-\sigma}}{(q;q)_j(q;q)_m(q;q)_{2n-m-\sigma}}.
\end{multline*}
After the shifts $n\to n-j$ and $m\to m-j$ this is
\begin{align*}
\text{LHS}\eqref{eqB}&=
\frac{1}{(q;q)_{\infty}}
\sum_{n=0}^{\infty}\sum_{m=0}^{2n-\sigma}
q^{kn^2+(\la-\sigma k)n}(1-q^{(k-\la)(2n-\sigma+1)})\\
&\qquad\qquad\qquad\times 
\frac{(q;q)_{2n-\sigma}}{(q;q)_m(q;q)_{2n-m-\sigma}}\:
{_2}\phi_1\biggl[\genfrac{}{}{0pt}{}
{q^{-m},q^{-(2n-m-\sigma)}}{q^{-(2n-\sigma)}};q,q\biggr] \\
&=\frac{1}{(q;q)_{\infty}}
\sum_{n=0}^{\infty}\sum_{m=0}^{2n-\sigma}
q^{kn^2+(\la-\sigma k)n}(1-q^{(k-\la)(2n-\sigma+1)}) \\
&=\frac{1}{(q;q)_{\infty}}
\sum_{n=-\infty}^{\infty}(2n-\sigma+1)
q^{kn^2+(\la-\sigma k)n}.
\end{align*}
Here the second equality follows by noting that the same
$_2\phi_1$ sum occurs in \eqref{Lphi} so that it
equates to \eqref{cv}. The last equality follows
from \eqref{last}.

\bibliographystyle{amsplain}

\end{document}